# Exploring the Level of Urbanization Based on Zipf's Scaling Exponent

Yanguang Chen

(Department of Geography, College of Urban and Environmental Sciences, Peking University, Beijing 100871, P.R. China. E-mail: chenyg@pku.edu.cn)

**Abstract**: The rank-size distribution of cities follows Zipf's law, and the Zipf scaling exponent often tends to a constant 1. This seems to be a general rule. However, a recent numerical experiment shows that there exists a contradiction between the Zipf exponent 1 and high urbanization level in a large population country. In this paper, mathematical modeling, computational analysis, and the method of proof by contradiction are employed to reveal the numerical relationships between urbanization level and Zipf scaling exponent. The main findings are as follows. (1) If Zipf scaling exponent equals 1, the urbanization rate of a large populous country can hardly exceed 50%. (2) If Zipf scaling exponent is less than 1, the urbanization level of large populous countries can exceeds 80%. A conclusion can be drawn that the Zipf exponent is the control parameter for the urbanization dynamics. In order to improve the urbanization level of large population countries, it is necessary to reduce the Zipf scaling exponent. Allometric growth law is employed to interpret the change of Zipf exponent, and scaling transform is employed to prove that different definitions of cities do no influence the above analytical conclusion essentially. This study provides a new way of looking at Zipf's law of city-size distribution and urbanization dynamics.

# 1. Introduction

Urbanization is a process of nonlinear dynamics in self-organized evolution of human settlements. Urbanization result in rank-size distribution at macro level and complex form and growth of cities at micro level. In intra-urban geography, urban form and growth can be modeled by fractal geometry (Batty and Longley, 1994; Frankhauser, 1994; Frankhauser, 1998). In interurban geography, the city size distribution can be described with Zipf's law (Zipf, 1949), which is also termed rank-size rule in the literature on cities (Carroll, 1982; Knox and Marston, 2009). The rank-size rule represents the pure Zipf distribution with scaling exponent equal to 1. This rule is simple, graceful, easy to understand, and can be found everywhere in urban studies (Batty and Longley, 1994; Gabaix, 1999a; Gabaix, 1999b; Jiang *et al*, 2015; Jiang and Jia, 2011; Krugman, 1996). For a long time, Zipf's distribution of city sizes were regarded as super-stable distribution (Buchanan, 2000; Knox and Marston, 2009; Madden, 1956; Zhou, 1995). The distribution models and the parameter values used to be regarded as depending on the development degree of a country (Berry, 1961; Roehner, 1991). However, in the real world, the Zipf scaling exponent of city size distribution in some countries such as China and the United States of America deviates significantly from 1 (Chen, 2012). New evidences show that the size distribution models of cities and the related scaling exponent are not always decided by the extent of economic development of a country. On the one hand, Zipf exponent value depends on the definition of city; on the other hand, Zipf exponent relies on population size of a country. The problem is that if one of the city's definitions is closely related to the level of urbanization, the logical contradiction can be deduced from the pure Zipf's law. Suppose that the population size of a country is more than 500 million, and the Zipf exponent is equal to 1, we can theoretically draw the following inference: the proportion of urban population in the total population of the country will hardly exceed 50%. In other word, based on pure Zipf size distribution of cities, it is hard to reach the level of urban majority for a large populous country. If the above inference is true, the urbanization dynamics of a large population country will inevitably lead to the deviation of the scaling exponent of rank size distribution of cities from 1.

The rank-size rule is one of the three basic laws in urban geography. Another two ones are distance-decay law and allometric growth law, respectively. The relationships between urbanization level and city size distribution is an old-fashioned problem in urban studies (Berry, 1961; Song and Zhang, 2002). The history of Zipf law in urban research can be traced back to Auerbach's law of population concentration (Auerbach, 1913). After that, Jefferson (1939) proposed the law of the primate city, which resulted in the concepts of the primate city and primate rule. Thus city size distribution were divided into two typical categories: rank-size type and primate type (Berry, 1961).

The two types of size distributions of cities can be distinguished from one another by the primacy ratio (Zhou, 1995), or by length ratio (Dimitrova and Ausloos, 2015). For a long time, a great many studies have been devoted to the field of city-size distributions, and many interesting findings emerged (Carroll, 1982; Gabaix and Ioannides, 2004; Zhou, 1995). In past 20 years, some scholars proposed universal rank-size law to go beyond simple power law (Ausloos and Cerqueti, 2016; Benguigui *et al*, 2007a; Benguigui *et al*, 2007b; Benguigui *et al*, 2011). Among various types of rank-size distribution models, Zipf law represents a self-organized critical state of complex spatial systems (Bak, 1996; Portugali, 2000). What is mathematical relationships between urbanization level and Zipf scaling exponent? What is physical mechanism causing the differentiation between the rank-size distribution and primate distribution? Based on mathematical derivation, computational analyses, and empirical evidence, this paper aims at the dynamical mechanism of urbanization resulting in variation of Zipf scaling exponent of rank-size distribution of cities. The rest parts are arranged as follows. In Section 2, the mathematical equations are derived to describe the relationships between urbanization level, city number, national population, and Zipf scaling exponent. In Section 3, a series of computational analyses based on related mathematical models are made to testify and complement the derived theoretical results in Section 2. In Section 4, several questions are responded and discussed, and finally, the discussion are concluded by summarizing the main points of this work.

## 2. Models

### 2.1 A formula of urbanization level based on pure Zipf's law

The well-known Zipf's law is the starting point of this theoretical exploration. A set of basic formula can be derived for computational and empirical analyses. Suppose that there are $N$ cities and towns in a geographical region, say, a country, and the rank-size distribution of these human settlements follows Zipf's law (Zipf, 1949). The sizes of cities and towns are measured by urban population. The common form of Zipf's law is often expressed as (Carroll, 1982)

$$P(k) = P_1 k^{-q}, \tag{1}$$

where $k$ denotes the rank of a city/town, $P(k)$ is the city population size of the $k$th city/town, $P_1$ refers to the size of the largest city, i.e., the primacy city, and $q$ can be termed Zipf scaling exponent.

Equation (1) is the well-known two-parameter Zipf model of city size distributions. It proved to be equivalent to Pareto's cumulative distribution function (CDF) (Chen *et al*, 1993). In the simplest case, the total urban population of a country is

$$U = \sum_{k=1}^{N} P(k) = P_1 \sum_{k=1}^{N} k^{-q}, \qquad (2)$$

where $U$ refers to the total urban population in a country. For the pure Zipf's distribution, the Zipf exponent $q=1$. The corresponding scaling exponent of Pareto's density distribution function (DDF) is $1/q+1=2$ (Manrubia and Zanette, 1998; Rozenfeld *et al*, 2011; Zanette and Manrubia, 1998). In this case, the two-parameter model is reduced to a one-parameter model, $P(k)=P_1/k$, which describes the inverse relationship between city rank and size (Zipf, 1949). According to Euler's formula for the sum of harmonic series, we have

$$\lim_{N \to \infty}(\sum_{k=1}^{N} \frac{1}{k} - \ln N) \to C = 0.577216\cdots, \qquad (3)$$

where $C=0.577216\ldots$ denotes one of Euler's constant (Havil, 2003). Thus, the total urban population is

$$U = \sum_{k=1}^{N} P(k) = P_1(C + \ln N). \qquad (4)$$

This is the discrete expression of total urban population. The level of urbanization of a country is defined as $L=U/P_T*100\%$, where $P_T$ represents entire population of a country, including urban population and rural population (United Nations, 1980a). So we have

$$L = \frac{U}{P_T} \times 100 = \frac{100 P_1(\ln(N) + C)}{P_T}, \qquad (5)$$

which represents the basic relation between urbanization level and Zipf's law of city-size distribution. The differential of $L$ with respect to $N$ is $dL/dN=100P_1/(P_T N)$, which suggests that if $N$ is large enough, new increasing towns will lead to very little increase of urbanization level. Equation (5) can be rewritten as

$$N = \exp(\frac{LP_T}{100P_1} - C) = N_0 \exp(\frac{LP_T}{100P_1}) = N_0 \exp(\frac{U}{P_1}), \qquad (6)$$

where the coefficient $N_0=\exp(-C)=\exp(-0.577216\ldots)=0.561459\ldots$. Equation (6) means that the number of cities increases exponentially as the level of urbanization goes up, and the size of the largest city is just the characteristic value of total population of cities and towns. Using equation (6), we can reveal the relationships between the level of urbanization, the population of the largest city, the number of cities and towns, the total urban population, and the total population of a country.

**2.2 Zipf's scaling exponent and total urban population**

The rank-size rule is regarded as a solution to the scaling functional equation of city size distribution, but the pure Zipf distribution does not reflect a strict scaling relation. For a scaling function, both its differential and integral results obey scaling law (Chen, 2015). However, the integral of pure Zipf's law does not satisfy the scaling relation. It is actually the intersection of inverse hyperbolic function and inverse power law. A power law is a special solution to the scaling functional equation. A hyperbolic function is not really a scaling relation, but it is always treated as linear scaling in literature. In mathematics, Zipf's distribution can be abstracted as simple problem of $p$-series. The $p$ is exactly the $q$ of Zipf's distribution. If $p=q=1$, the $p$-series will change to the harmonic sequence. The changing regularity of a sequence can be judged through generalized integral. If one is familiar with advanced mathematics, he can easily understand the following analysis. For the case $q=1$ in equation (1), we have

$$T = \int_{k=1}^{N} P(k)\mathrm{d}k = P_1 \int_{k=1}^{N} \frac{1}{k} \mathrm{d}k = P_1 \left[\ln k\right]_1^N = P_1 \ln N , \qquad (7)$$

where $T$ denotes the total urban population in a region. This is the continuous expression of urban total population. The integral result is a logarithmic function rather than a power function. This suggests that the pure Zipf's law is a semi-scaling relation. The population size of the largest city, $P_1$, is the characteristic value of the total urban population, $T$. The difference between the sum of discrete sequences, $U$, and the integration of continuous variable, $T$, is

$$U - T = P_1(\sum_{k=1}^{N} \frac{1}{k} - \int_{k=1}^{N} \frac{1}{k} \mathrm{d}k) = CP_1 . \qquad (8)$$

This implies $U=T+CP_1$. The error comes from approximate processing and does not reflect the essence of the problem, and thus does not affect the analytical conclusion.

The reasoning results based on discrete variables are more accurate, but the reasoning processes are often more difficult. The analytical process based on continuous variables is simpler than that based on discrete variables. Generally speaking, theoretical studies are always based on continuous variable, and calculus is utilized, while application studies are usually based on discrete variables, and the methods of difference are adopted. If the exponent $q \neq 1$, the Zipf model returns to its common form. In this instance, we have

$$T = \int_{k=1}^{N} P(k) \mathrm{d}k = P_1 \int_{k=1}^{N} k^{-q} \mathrm{d}k = P_1 \left[ \frac{k^{1-q}}{1-q} \right]_1^N. \tag{9}$$

Suppose that the city number, $N$, is large enough. The $N$ value will be infinite in theory. Then the total urban population is

$$T = P_1 \lim_{N \to \infty} \left[ \frac{k^{1-q}}{1-q} \right]_1^N = P_1 \lim_{N \to \infty} \left( \frac{N^{1-q}}{1-q} - \frac{1}{1-q} \right) = \begin{cases} \dfrac{P_1}{q-1}, & q > 1 \\ \infty, & q < 1 \end{cases}. \tag{10}$$

This implies that different $q$ values lead to different results. If $q>1$, the total urban population is limited, while if $q \leq 1$, the total urban population is not limited. For $q>1$, the level of urbanization is

$$L = \frac{100T}{P_T} = \frac{100P_1}{P_T} \lim_{N \to \infty} \left( \frac{N^{1-q}}{1-q} - \frac{1}{1-q} \right) = \frac{100P_1}{(q-1)P_T}, \tag{11}$$

which suggests that there is no significant relationships between urbanization level and city number. Consequently, it is impossible to promote the level of urbanization of a large populous country by adding more cities and towns, unless the property of rank-size distribution is changed. In fact, the number of cities and towns for $q>1$ can be derived from equation (11) as below:

$$N = \left[ \frac{(1-q)P_T L}{100 P_1} + 1 \right]^{1/(1-q)}, \tag{12}$$

which is valid only for the case $L<50\%$. If urbanization level $L \geq 50\%$, then city number $N$ will go to infinity. This is absurd. The case of exponent $q>1$ is suitable for the smaller countries, and the population size of the largest city determine the urbanization level. In extreme cases, the law of primate city will replace the rank-size law to dominate the city size distribution of a country. Maybe to a degree equation (11) can account for the law of primate city, which was proposed by Jefferson (1939). In contrast, for $q<1$, we have

$$L = \frac{100T}{P_T} = \frac{100(N^{1-q}-1)P_1}{(1-q)P_T}. \tag{13}$$

If city number $N$ is large enough, then an approximate power law can be derived as

$$L = \frac{100T}{P_T} = \lim_{N \to \infty} \frac{100(N^{1-q}-1)P_1}{(1-q)P_T} = \frac{100 P_1 N^{1-q}}{(1-q)P_T}. \tag{14}$$

This formula is suitable for the largely populous nations such as China and India. In this case, the level of urbanization depends heavily on city number and the population size of the largest city. Clearly, if the national size $P_T$ is large enough, the number of cities and towns can be expressed as

$$N = [\frac{(1-q)P_T L}{100 P_1} + 1]^{1/(1-q)} \approx [\frac{(1-q)P_T L}{100 P_1}]^{1/(1-q)}. \tag{15}$$

which suggests that a limited number of urban settlements in a region will allow high level of urbanization (e.g., $L$=80%). As for $q$=1, the level of urbanization can be formulated by equation (5), which can be approximated to $(100P_1/P_T)\ln N$. For given population size of the largest city, equation (13) suggests an approximate power law relation between city number $N$ and total urban population $T$, that is

$$T = \frac{P_1}{1-q} N^{1-q}, \tag{16}$$

which differs the exponential relation, equation (6). This suggests an allometric relation between city number and total urban population. For given city number $N$, total urban population is proportional to the population size of the largest city. For a country, equation (16) can be used to describe the cumulative distribution of city sizes based on general Zipf distribution; for different regions (e.g., a system of provinces, a system of states) in a country, equation (16) can be used to describe the aggregate relation between city number and total urban population.

## 2.3 The definitions of urban population and city number

Above mathematical relations involve the variable of urban population and total urban population. It is necessary to explain the two concepts. Urban population depends on the identification of urban boundaries (Davis, 1978; Zhou, 1995), while total urban population depends on both the identification of urban boundaries and determination of city number in a country (Chen, 2003). Unfortunately, both urban boundary and city number can only be reasonably defined, but cannot be objectively determined. Urban form bear fractal nature (Batty and Longley, 1994; Frankhauser, 1994), thus it is impossible to identify urban boundaries exactly (Chen *et al*, 2019). In urban geography, there exist three core concepts of cities: city proper (CP), urbanized area (UA), and metropolitan areas (MA) (Chen, 2008; Davis, 1978; Knox and Marston, 2009; Rubenstein, 1999; Zhou, 1995). The second concept is sometimes substituted with urban agglomerations (UA).

Different city concepts correspond to different urban boundaries, and different urban boundaries indicate different urban population sizes (Davis, 1978). On the other hand, total population of cities depends on city number. The rank-size rule implies scaling property of city size distributions. Scaling suggests no typical size can be found for cities (Buchanan, 2000; Chen, 2003). As a results, different countries have different cut-off criterions of population size for urban settlements (Knox and Marston, 2009; United Nations, 1980a; United nations, 1980b; Zhou, 1995). What is more, the minimum population size of cities in a country varies with city development (Chen, 2003). In this case, the number of cities and towns in a country cannot be determined objectively.

In short, due to fractal feature and scaling nature of cities, urban population and city number cannot be objectively determined. Fortunately, urban population density follows the distance-decay law (Chen, 2015). Therefore, in practice, urban boundaries can be defined by the city clustering algorithm (CCA) proposed by Rozenfeld *et al* (2008). On the other hand, the scaling range of the rank-size distribution of cities in reality is limited (Chen, 2016). Thus, city number can be determined according to the lower limit of scaling range of city-size distribution. If urban boundaries are identified, urban population $P(k)$ can be effectively defined. If the city number $N$ and urban population $P(k)$ are determined, the total urban population $U$ can be scientifically determined. No matter how to define the urban population and the total population, as long as the size distribution of cities in a populous country follows Zipf's law, the conclusion caused by the above mathematical reasoning will not change. This judgment can be proved by scaling transformation (see Section 3).

The level of urbanization is defined by the ratio of total urban population to the total population of a country. It is easy to determine the total population in a geographical region. However, the definition of the total urban population relies heavily on the determination of number of cities and towns. According to the scaling nature of Zipf's distribution of cities, the city number is a variable rather than a constant in theory. Due to the scaling property of urban form and city-size distributions, we can utilize the method of proof by contradiction (reduction to absurdity) to prove the relation between the scaling exponent of Zipf distribution $q$ and urbanization level $L$. If the Zipf scaling exponent $q \geq 1$, even if the number of cities and towns in a large population country is infinite ($N \to \infty$), the level of urbanization cannot reach 80%. In contrast, if $q \geq 1$, a finite number of cities and towns in a country ($N \ll P_1$) allows the urbanization level to exceed 80%. Please note that, in above analytical process, the value of variable $N$ depends on the value of parameter $q$. For the case of $q \geq 1$, if an infinite number of cities and towns cannot accommodate 80% of the national population, how can a finite number of cities and towns allow 80% of the country's population? The general conclusion can be drawn from the infinite numbers ($N \rightarrow \infty$), while special conclusion can be drawn from the finite number. Any readers can readily test this theoretical inference by a series of simple computational analyses.

## 3. Computational and empirical analyses

### 3.1 Computational results and analyses

Mathematical experiment based on the above-shown formulae can be employed to explore the relationships between urbanization and Zipf's law. Mathematical experiment is in fact a type of computational method. Based on certain scientific principle, a series of numerical computation can be conducted with basic assumptions and mathematical models (Chen, 2012). New findings can be made by comparing the computational results with mathematical reasoning results and observational phenomena. The typical formulae presented in Section 2 will be utilized to make computational analysis. The computational process consists of four parts. (1) Computing urbanization levels for populous nations based on given parameters. (2) Computing city numbers for populous nations based on given parameters. (3) Computing urbanization levels for different sizes of nations based on given parameters. (4) Computing urbanization levels for populous nations based on given parameters in terms of general Zipf's law. The first three parts are based on pure Zipf's law, and the final part is based on general Zipf's law. Based on the computational results, the reduction to absurdity can be employed to reveal the contradiction between the pure Zipf law with high level of urbanization ($L>70\%$). Let us make computation and analyses step by step.

**Mathematical experiment 1: computing urbanization levels for populous nations based on given parameters.** Based on pure Zipf's rank-size distribution of cities, the levels of urbanization are derived for a country with a large population. Suppose a country has a population of 1.4 billion ($P_T=1.4*10^9$). Two approaches can be used to predict the levels of urbanization based on different numbers of cities and towns ($N$) and different population sizes of the largest city ($P_1$). One is the brute force method based on equation (1), and the other is the simple method based on equation (5). The results show that for such a large population, the level of urbanization is limited by 50% (Table 1). If the population of the largest city is twenty million ($P_1=2*10^7$), then 500,000 cities and towns can only accommodate about 19.57% of the population ($L<20\%$). In this case ($N$=500,000), if the urbanization rate is expected to exceed 50%, the population size of the largest city should reach more than 50 million. The size of the largest city is over large. There will be about 60 cities with a population of more than 1 million. This is not normal urban population size and number of cities. A conclusion can be drawn that, based on normal size of the largest city and city number, the level of urbanization in a country with 1.4 billion people can hardly exceed 50%.

**Table 1 The levels of urbanization based on different city numbers and different population sizes of the largest city**

| City number, $N$ | Level of urbanization, $L$ (%) | | | | | |
|---|---|---|---|---|---|---|
| | $P_1$=1000 | $P_1$=2000 | $P_1$=3000 | $P_1$=4000 | $P_1$=5000 | $P_1$=6000 |
| **1000** | 5.3468 | 10.6935 | 16.0403 | 21.3871 | 26.7338 | 32.0806 |
| **5000** | 6.4961 | 12.9922 | 19.4882 | 25.9843 | 32.4804 | 38.9765 |
| **10000** | 6.9911 | 13.9823 | 20.9734 | 27.9646 | 34.9557 | 41.9469 |
| **50000** | 8.1407 | 16.2814 | 24.4222 | 32.5629 | 40.7036 | 48.8443 |
| **100000** | 8.6358 | 17.2716 | 25.9075 | 34.5433 | 43.1791 | 51.8149 |
| **200000** | 9.1309 | 18.2618 | 27.3928 | 36.5237 | 45.6546 | 54.7855 |
| **300000** | 9.4205 | 18.8411 | 28.2616 | 37.6822 | 47.1027 | 56.5232 |
| **400000** | 9.6260 | 19.2521 | 28.8781 | 38.5041 | 48.1301 | 57.7562 |
| **500000** | 9.7854 | 19.5708 | 29.3562 | 39.1417 | 48.9271 | 58.7125 |

**Note**: (1) The symbols are as follows: $N$—number of cities and towns; $L$—level of urbanization (%); $P_1$—urban population of the largest city. (2) The national total population is assumed to be $P_T=14*10^8$, i.e., 1 billion 400 million. (3) The population unit of $P_1$ is 10 thousand.

**Mathematical experiment 2: computing city numbers for populous nations based on given parameters.** For given level of urbanization, the number of cities and towns can be worked out. Two approaches can be employed to predict the city number based on different levels of urbanization ($L$) and different population sizes of the largest city ($P_1$). One is the brute force method by using equation (1), and the other is the simple method by means of equation (6). For a country bearing a population of 1.4 billion, the results are tabulated as below (Table 2).The high level of urbanization corresponds to the abnormal number of towns. Suppose the population size of the largest city is twenty million ($P_1=2*10^7$). If the urbanization level reach 80%, it will need $1.1744*10^{24}$ cities and towns. The number of cities and towns is $8.39*10^{18}$ times the national population. The small towns increased afterwards can accommodate less than one person. The result is absurd. The reason is that the standard Zipf distribution is not suitable for a country with a very large population such as China and India. So, is the pure Zipf law invalid? Of course not. For countries with small populations, the law works. An absurd result is that for a large country of 14 people, if the level of urbanization reaches 80%, the number of cities and towns is greater than that of people. A conclusion can be reached by the method of reduction to absurdity that the pure Zipf law is not suitable for the country with 1.4 billion people such as China.

**Table 2 The city number based on different levels of urbanization and different population sizes of the largest city**

| $L$ (%) | Number of total cities and towns, $N$ | | | | | |
|---|---|---|---|---|---|---|
| | $P_1$=1000 | $P_1$=2000 | $P_1$=3000 | $P_1$=4000 | $P_1$=5000 | $P_1$=6000 |
| 10 | 675213 | 616 | 60 | 19 | 9 | 6 |
| 20 | 8.1201E+11 | 675213 | 6349 | 616 | 152 | 60 |
| 30 | 9.7653E+17 | 7.4046E+08 | 675213 | 20390 | 2497 | 616 |
| 40 | 1.1744E+24 | 8.1201E+11 | 7.1804E+07 | 675213 | 41060 | 6349 |
| 50 | 1.4123E+30 | 8.9048E+14 | 7.6358E+09 | 2.2360E+07 | 675213 | 65477 |
| 60 | 1.6985E+36 | 9.7653E+17 | 8.1201E+11 | 7.4046E+08 | 1.1104.E+07 | 6.7521.E+05 |
| 70 | 2.0426E+42 | 1.0709E+21 | 8.6352E+13 | 2.4521E+10 | 1.8260.E+08 | 6.9630.E+06 |
| 80 | 2.4564E+48 | 1.1744E+24 | 9.1829E+15 | 8.1201E+11 | 3.0027.E+09 | 7.1804.E+07 |
| 90 | 2.9541E+54 | 1.2879E+27 | 9.7653E+17 | 2.6890E+13 | 4.9379.E+10 | 7.4046.E+08 |
| 100 | 3.5526E+60 | 1.4123E+30 | 1.0385E+20 | 8.9048E+14 | 8.1201.E+11 | 7.6358.E+09 |

**Note**: (1) The symbols are as follows: $L$—level of urbanization (%); $N$—number of cities and towns; $P_1$—urban population of the largest city. (2) The national total population is assumed to be $P_T$=14*10$^8$. (3) The population unit of $P_1$ is 10 thousand.

**Mathematical experiment 3: computing urbanization levels for different sizes of nations based on given parameters.** Let's consider the countries with smaller populations. Suppose that the population of the countries ranges from 100 million to 500 million. Using equation (1) or equation (5), we can calculate the levels of urbanization based on different numbers of cities and towns and national population sizes (Table 3). For a country with a population of 100 million, if the size of the largest city is $P_1$= 10 million, 2000 cities and towns can accommodate 81.78% of the population. In other words, the level of urbanization can easily exceed 80% if its level of industrialization allows. On the other hand, for countries with a population of 100 million, the population size of the largest cities should not exceed 10 million. If the largest city size reaches 20 million, either the total number of cities is small, or the rank-size distribution of cities breaks. In

light of equation (6), based on pure Zipf distribution, about 83 cities can accommodate the whole population of the country. In this case, if the national territory is not large enough, its city size may follow the primate distribution. The largest city is very large, but the second largest city is far smaller than the largest one. In other words, the primate degree is significantly greater than 2, say, $P_1/P_2>3$, where $P_1$ and $P_2$ denote the population sizes of the first and second largest cities. A conclusion is that, for the countries with small population, the pure Zipf's law works. However, if the largest city become too large, the rank-size distribution will change to the primate distribution.

**Table 3 The level of urbanization based on different city numbers, different population sizes of the largest city, and different national total population**

| City number, $N$ | Level of urbanization, $L$ (%) | | | | | | | | | |
|---|---|---|---|---|---|---|---|---|---|---|
| | $P_1=1000$ | | | | | $P_1=2000$ | | | | |
| | $P_T=10000$ | $P_T=20000$ | $P_T=30000$ | $P_T=40000$ | $P_T=50000$ | $P_T=10000$ | $P_T=20000$ | $P_T=30000$ | $P_T=40000$ | $P_T=50000$ |
| 500 | 67.9282 | 33.9641 | 22.6427 | 16.9821 | 13.5856 | 135.8565 | 67.9282 | 45.2855 | 33.9641 | 27.1713 |
| 1000 | 74.8547 | 37.4274 | 24.9516 | 18.7137 | 14.9709 | 149.7094 | 74.8547 | 49.9031 | 37.4274 | 29.9419 |
| 2000 | 81.7837 | 40.8918 | 27.2612 | 20.4459 | 16.3567 | 163.5674 | 81.7837 | 54.5225 | 40.8918 | 32.7135 |
| 3000 | 85.8375 | 42.9187 | 28.6125 | 21.4594 | 17.1675 | 171.6750 | 85.8375 | 57.2250 | 42.9187 | 34.3350 |
| 4000 | 88.7139 | 44.3570 | 29.5713 | 22.1785 | 17.7428 | 177.4278 | 88.7139 | 59.1426 | 44.3570 | 35.4856 |
| 5000 | 90.9451 | 45.4725 | 30.3150 | 22.7363 | 18.1890 | 181.8902 | 90.9451 | 60.6301 | 45.4725 | 36.3780 |
| 6000 | 92.7681 | 46.3841 | 30.9227 | 23.1920 | 18.5536 | 185.5363 | 92.7681 | 61.8454 | 46.3841 | 37.1073 |
| 7000 | 94.3095 | 47.1548 | 31.4365 | 23.5774 | 18.8619 | 188.6191 | 94.3095 | 62.8730 | 47.1548 | 37.7238 |
| 8000 | 95.6447 | 47.8224 | 31.8816 | 23.9112 | 19.1289 | 191.2895 | 95.6447 | 63.7632 | 47.8224 | 38.2579 |
| 9000 | 96.8225 | 48.4113 | 32.2742 | 24.2056 | 19.3645 | 193.6450 | 96.8225 | 64.5483 | 48.4113 | 38.7290 |
| 10000 | 97.8761 | 48.9380 | 32.6254 | 24.4690 | 19.5752 | 195.7521 | 97.8761 | 65.2507 | 48.9380 | 39.1504 |

**Note**: (1) The symbols are as follows: $N$—number of cities and towns; $L$—level of urbanization (%); $P_1$—urban population of the largest city; $P_T$—national total population. (2) The level of urbanization is not greater than 100, that is, $L \leq 100\%$. Otherwise, the number is abnormal. (3) The population unit of $P_1$ and $P_T$ is 10 thousand.

**Mathematical experiment 4: computing urbanization levels for populous nations based on given parameters in terms of general Zipf's law.** For given national population in a country, the level of urbanization depends on city number, the population size of largest city, and Zipf scaling exponent. This can be reflected by equations (5) and (13). Equation (5) is available for the case of $q=1$, while equation (13) is available for the case of $q \neq 1$. According to above analysis, the first situation is not suitable for a big country like China. Let us look at the second situation. Suppose a country has 1.4 billion people. The population of the largest city is considered to be two cases: one is 10 million, and the other, 20 million. Mathematical experiments show that with the increase of

the largest city size and the decrease of Zipf scaling exponent, the level of urbanization rises rapidly (Table 4). The impact of the number of cities and towns on the level of urbanization is not particularly significant. If $P_1$=20 million and $q$=0.8, then 300 thousand cities and towns can hold about 82.64% of the national population. A conclusion is that, for a country with a large population like China, urbanization dynamics will lead to the rank-size distribution of cities with Zipf exponent less than 1.

**Table 4 The level of urbanization based on different city numbers, different population sizes of the largest city, and different Zipf scaling exponent values**

| City number, $N$ | Level of urbanization, $L$ (%) | | | | | | | | | |
|---|---|---|---|---|---|---|---|---|---|---|
| | $P_1$=1000 | | | | | $P_1$=2000 | | | | |
| | $q$=0.8 | $q$=0.9 | $q$=1 | $q$=1.1 | $q$=1.2 | $q$=0.8 | $q$=0.9 | $q$=1 | $q$=1.1 | $q$=1.2 |
| 1000 | 11.0499 | 7.5168 | 5.3468 | 3.9806 | 3.0970 | 22.0997 | 15.0336 | 10.6935 | 7.9612 | 6.1939 |
| 5000 | 16.4479 | 10.0049 | 6.4961 | 4.5126 | 3.3438 | 32.8957 | 20.0098 | 12.9922 | 9.0253 | 6.6876 |
| 10000 | 19.3647 | 11.2063 | 6.9911 | 4.7167 | 3.4280 | 38.7295 | 22.4127 | 13.9823 | 9.4334 | 6.8559 |
| 50000 | 27.9215 | 14.3393 | 8.1407 | 5.1394 | 3.5837 | 55.8430 | 28.6786 | 16.2814 | 10.2789 | 7.1675 |
| 100000 | 32.5447 | 15.8519 | 8.6358 | 5.3016 | 3.6368 | 65.0893 | 31.7038 | 17.2716 | 10.6031 | 7.2737 |
| 200000 | 37.8553 | 17.4731 | 9.1309 | 5.4528 | 3.6831 | 75.7106 | 34.9462 | 18.2618 | 10.9056 | 7.3662 |
| 300000 | 41.3207 | 18.4749 | 9.4205 | 5.5366 | 3.7073 | 82.6415 | 36.9497 | 18.8411 | 11.0731 | 7.4146 |
| 400000 | 43.9556 | 19.2107 | 9.6260 | 5.5939 | 3.7233 | 87.9112 | 38.4213 | 19.2521 | 11.1879 | 7.4466 |
| 500000 | 45.7046 | 19.3891 | 9.7854 | 5.2199 | 3.3126 | 91.4093 | 38.7782 | 19.5708 | 10.4398 | 6.6252 |

**Note**: (1) The symbols are as follows: $N$—number of cities and towns; $L$—level of urbanization (%); $P_1$—urban population of the largest city. (2) The national total population is assumed to be $P_T$=14*10$^8$. (3) The population unit of $P_1$ is 10 thousand.

### 3.2 Empirical analysis of Chinese cities

China is famous for its large population, so Chinese mainland cities can be employed to testify the results of theoretical derivation and calculation analyses. In fact, the problems explored in this paper are just caused by the theoretical dilemma of China's urbanization and city size distribution analysis. Whether the size distribution of Chinese cities follows Zipf's law is a controversial issue. Some scholars examined China's urban rank-size patterns by means of Zipf's law (e.g., Chen *et al*, 1993; Chen and Zhou, 2008; Gan *et al*, 2006; Gangopadhyay and Basu, 2009; Peng, 2010; Schaffar and Dimou, 2012; Xu and Zhu, 2009; Ye and Xie, 2012; Zhou, 1995), and others regarded China's city size distribution as non-Zipf's distribution (e.g., Anderson and Ge, 2005; Benguigui and Blumenfeld-Lieberthal, 2007a; Benguigui and Blumenfeld-Lieberthal, 2007b). In fact, the rank-size distribution of Chinese cities can be described by the three-parameter Zipf's law (general model) (Chen, 2016). The three-parameter model is often termed Zipf-Mandelbrot-Pareto model in literature (Ausloos, 2014; Mandelbrot, 1982). If we use the one-parameter Zipf's law (pure model) or the two-parameter Zipf's law (common model) to describe China's city size distribution, the results seem to be ambiguous and thus inconclusive. The reason lies in that China's urban area and size threshold are of great administrative significance. Urban census data only include officially approved cities. A large number of important cities lack census data because they are not included in the official city list.

By using the data of four times population censuses, we can thoroughly investigate the basic characteristics and changing trends of rank size distribution of cities in mainland China. Four datasets of city sizes include the observations of the third census (1982), the fourth census (1990), the fifth census (2000), and the sixth census (2010). Two aspects of explanation are as follows. Frist, in the year following the census, local governments published preliminary results through census bulletins. However, these results often have errors that cannot be ignored. The National Bureau of Statistics of China used the unified standard to calibrate the census data of all cities. In this work, the adjusted data are adopted. Second, the population size of cities in China include total population of each city (nonagricultural population + agricultural population), population in urban areas (city population + township population), and registered residence population. In this study, the city population is adopted (Table 5). Draw the data points on a double logarithmic plot for rank-size relation. If the points form a straight line, they can be regarded as following Zipf's law.

**Table 5 The basic data of Chinese cities from four times of population census**

| Year | Total population $P_T$ | Urban population $U$ | City number $N$ | Largest city size $P_1$ | Urbanization level $L$ |
|---|---|---|---|---|---|
| **1982** | 100818.0000 | 21082.000 | 238 | 6320829 | 20.9109 |
| **1990** | 113368.0000 | 29971.000 | 460 | 7469509 | 26.4369 |
| **2000** | 126583.0000 | 45844.000 | 666 | 12720701 | 36.2166 |
| **2010** | 133971.9546 | 66557.000 | 654 | 17640842 | 49.6796 |

**Note**: All the population data represent the finally adjusted city population rather than the preliminarily published total population.

The results show that not all the data points in each year take on straight trend on the rank-size log-log plot. The straight segments can be regarded as a scaling range of Zipf distribution. Fitting the data points within the scaling range to common Zipf's model, we can estimate the model parameters by the regression coefficients. The intercept represents the theoretical value of the population size of the largest city, and the slope represents Zipf scaling exponent (Table 6). The $P_1$ value in the model is very large, and the largest city in reality is far from reaching this size. This shows that the urbanization dynamics needs the largest city to increase to such as size, but the actual environmental, economic, and technical conditions cannot support it. All the Zipf scaling exponent values are less than 1, but from 1990 to 2010, the exponent values went up and up. Of course, this is only an approximate estimation. China's city size distributions were not well fitted by the conventional Zipf model (Chen, 2016).

Since the reform and opening up in 1980s, the speed of urbanization in China has gradually accelerated. During this period, a series of research reports are consistent with the theoretical analysis of this paper. For example, Zipf exponent values of Chinese city-size distribution became smaller (Chen *et al*, 1993; Gan *et al*, 2006), or small cities grow faster than big cities (Schaffar and Dimou, 2012; Xu and Zhu, 2009). Another related evidence comes from India. The Zipf exponent of the largest 1188 cities in 2011 is about $q$=1.1627. This exponent value is close to or even slightly greater than 1. According to Indian census, the level of urbanization of India in 2011 is about 31.1608%. The urbanization ratio is lower than the level of urban majority, $L$=50%.

Table 6 The parameter values of two-parameter Zipf modeling for Chinese cities and related numbers and statistics

| Year | Scaling range $N^*$ | Coefficient $P_1^*$ | Zipf scaling exponent $q$ | Goodness of fit $R^2$ |
|---|---|---|---|---|
| **1982** | 200 | 19080947.5526 | 0.9518 | 0.9504 |
| **1990** | 400 | 17271080.7380 | 0.8705 | 0.9819 |
| **2000** | 600 | 34743311.2477 | 0.8955 | 0.9865 |
| **2010** | 600 | 56265884.6457 | 0.9453 | 0.9842 |

**Note**: (1) The scaling range means that the $N^*$ largest cities approximately form a straight line on a log-log plot. (2) The model's proportionality coefficient $P_1^*$ represents the population size of the largest city in theory.

## 4. Discussion

The mathematical reasoning and computational analyses indicate that urbanization dynamics may change the scaling character of city-size distribution. The pure Zipf distribution is suitable for the countries with smaller population, but not suitable for the countries with large population. For countries with a population of no more than 100 million, the Zipf scaling exponent of city size distribution is always close to 1. However, if the population of a country is more than 500 million, the Zipf scaling exponent of rank-size distribution of cities will be less than 1, otherwise the level of urbanization will hardly exceed 50%. Zipf scaling exponent of city size distribution can be equal to or even greater than 1 for countries with medium and small populations ($q \geq 1$). For large population countries, the scaling exponent of city size distribution will so decrease due to urbanization that its value is less than 1 ($q<1$). In short, countries of different population sizes may exhibit different characteristics of size distribution of cities (Table 7). For $q \geq 1$, if the gap between $P_1$ and $P_T$ is smaller, the rank-size distribution may change into the primate distribution. In particular, for the case $q>1$, urbanization is mainly determined by the largest city, and if the largest city attract too many population, the primate distribution will probably come into being (Figure 1).

**Table 7 The relationships between total urban population, the number of cities and towns, and the Zipf's scaling exponent**

| Zipf exponent | Total urban population ($T$ or $U$) | Number of cities and towns ($N$) | Urbanization level ($L$) | Country size ($P_T$) | Urbanization percent |
|---|---|---|---|---|---|
| $q>1$ | $\dfrac{P_1(N^{1-q}-1)}{1-q}$ | $[\dfrac{(1-q)P_T L}{100 P_1}+1]^{1/(1-q)}$ | $\dfrac{100 P_1}{(q-1)P_T}$ | Small | Low |
| $q=1$ | $P_1(C+\ln N)$ | $\exp(\dfrac{L P_T}{100 P_1}-C)$ | $\dfrac{100 P_1(\ln(N)+C)}{P_T}$ | Small or medium-sized | High or low |
| $q<1$ | $\dfrac{P_1(N^{1-q}-1)}{1-q}$ | $[\dfrac{(1-q)P_T L}{100 P_1}+1]^{1/(1-q)}$ | $\dfrac{100 P_1 N^{1-q}}{(1-q)P_T}$ | Large | High |

The question is, does the definition and number of cities in a country affect the previous analysis and inferences? Scaling analysis can be used to give an answer to the question. As indicated above, equations (5), (11), and (13) involve two parameters and two variables. The two parameters are the scaling exponent, $q$, and the proportionality coefficient of Zipf's model, $P_1$. The two variables are the number of cities and towns, $N$, and the total national population, $P_T$. For given national population $P_T$ and Zipf scaling exponent $q$, the influence factors for urbanization level are city number, $N$, and the proportionality coefficient, $P_1$, indicating the size of the largest city. The size of the largest city relied heavily on the definition of cities in a country. Next, I will give two proofs by means of the ideas from scaling. First of all, the different definitions of cities do not influence the main conclusions of this work. It can be mathematically proved that as long as the Zipf pattern of city size distribution remains unchanged, the analytical conclusion will not change due to different urban boundaries. Zipf's law is in fact a scaling law of the rank-size distribution. Applying scaling transform to equation (1) yields

$$P(\zeta k) = P_1(\zeta k)^{-q} = \zeta^{-q} P_1 k^{-q} = \lambda P(k), \tag{17}$$

where $\zeta$ refers to a scale factor, and $\lambda=\zeta^{-q}$ denotes the eigenvalues of the scaling transform. Suppose that the rank-size distribution follows Zipf's law. Equation (17) implies rescaling city size according to different city definitions. Changing the definition of a city is equivalent to changing the scale factor in equation (17). The result is only equivalent to changing the proportional coefficient $P_1$. Assuming that new proportionality coefficient is $P_1^*$, we have

$$P_1^* = \lambda P_1, \tag{18}$$

Thus equation (4) will be replaced by

$$U^* = \lambda U = \lambda P_1 (C + \ln N). \tag{19}$$

So, equations (5), (11), and (13) can be re-expressed as

$$L = \frac{U^*}{P_T} \times 100 = \begin{cases} 100\lambda P_1 / ((q-1)P_T), & q > 1 \\ 100\lambda P_1 (\ln(N) + C) / P_T, & q = 1, \\ 100\lambda P_1 N^{1-q} / ((1-q)P), & q < 1 \end{cases} \tag{20}$$

in which the variation range and amplitude of eigenvalue $\lambda$ are limited. In Tables 1, 2, 3, and 4, changing the largest urban population size $P_1$ is equivalent to changing the eigenvalue $\lambda$, which is equivalent to changing the definition of a city. In terms of the above computational results, changing the definitions of cities do not influence the analytical conclusions.

Further, different standards for determining the number of cities will not affect the analysis and inferences of this paper. As indicated above, Zipf's distribution implies scaling. This suggests that it is impossible to find the objective size threshold for the definition of cities in a country (Chen, 2003). In fact, the use of proof by contradiction in this study is based on the uncertain number of cities. For example, according to equation (6), for a large country with a population of 1.4 billion, assuming that the largest city size is 50 million, based on the pure Zipf law ($q=1$), if the urbanization level reaches 80%, 3 billion cities and towns are needed (Table 2). A country of 1.4 billion people needs 3 billion cities and towns. This is a paradox. It is this absurdity that means that pure Zipf's law is not suitable for large population countries like China and India.

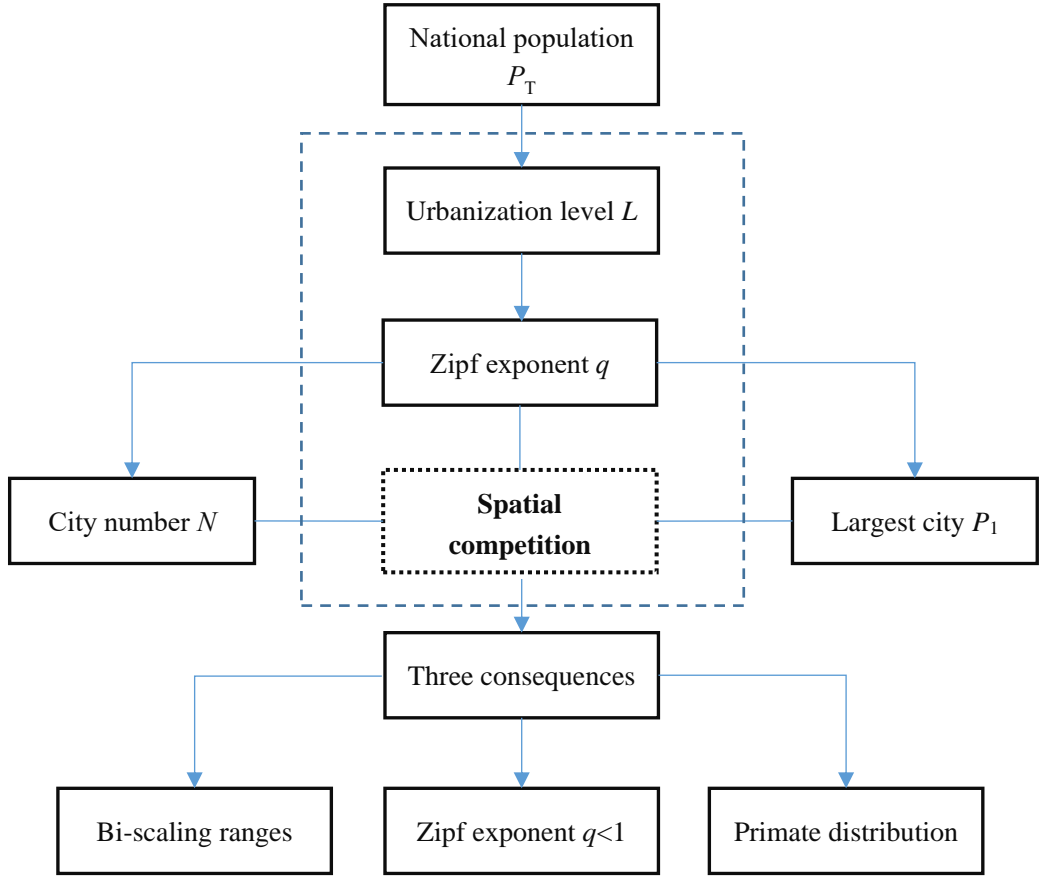

**Figure 1 The antecedents and consequences of the macro-level correlation between city size distributions and urbanization dynamics**

**Note**: For given national total population $P_T$ and Zipf scaling exponent $q$, the level of urbanization $L$ depends heavily on the number of cities and towns $N$ and the population of the largest city $P_1$. The interaction between different elements leads to three possible results: (1) The Zipf exponent reduce to $q<1$; (2) The scaling range breaks into two parts; (3) The rank-size distribution evolve into primate distribution.

The mathematical model of a system reflects the system's structure at macro level, while the model parameters reflect the distribution and relationships of elements at micro level. The Zipf scaling exponent is a parameter of the rank-size distribution model. Sometimes, the model may change in structure. The rank-size distribution will turn into the primate distribution. The rank-size distribution is based on Zipf's law (Zipf, 1949), which came from the *law of population concentration* (Auerbach, 1913). In contrast, the primate distribution is based on the *law of the* *primate city* (Jefferson, 1939). Based on Zipf's law and primate city law, city size distributions were divided into three categories: rank size distribution, primate distribution, and intermediate distribution (Berry, 1961). For a long time, geographers and economists have tried to explain the context of the different types of size distribution of cities. The rank-size distribution and primate distribution used to be regarded as two extreme types. According to the current evidences, that may not be the case. The primate distribution may be a local disturbance phenomenon of the rank-size distribution (Table 8). The above mathematical derivations and computational analyses provide new way for understanding this traditional academic problem. In fact, it is impossible for a large population country to have the primate size distribution of cities. The necessary conditions for the primate distribution of city sizes are as follows. First, small national territory and small population. Second, urbanization causes rapid growth of the primacy city. Third, the primacy city transcends national boundaries and becomes a member of international city network. Urbanization leads to the population expansion of the cities with the best conditions (generally the national capital). If the all cities grow according to Zipf's law, the country has neither enough population supplement nor enough resource support. In this way, the primacy cities form a *shadow effect* in the hierarchy. Just as a tall tree deprives the surrounding plants of sunlight and thus inhibit the growth of the nearby plants, the second, third… or even tenth largest cities are covered and limited by the first largest cities. An inevitable result is the primate distribution of size of cities in a country. By the way, the concept of *hierarchical shadow effect* differs from that of *spatial shadow effect* (Chen, 2011; Chen, 2014a; Evans, 1985).

Table 8 A comparison between the rank-size distribution and primate distribution of cities

| Category | Population law | Distribution rule | Country type | Rule type | Origin |
|---|---|---|---|---|---|
| **Rank size** | Law of population concentration | Zipf's law, rank-size rule | Countries of any size | Global rule | The most probable distribution |
| **Primate** | Law of the primate city | Primate rule | Smaller countries | Local rule | Hierarchical shadow effect |

In theory, Zipf's size distribution of cities is associated with urban allometric growth. To explain the Zipf scaling exponent, a *q*-equation of urban hierarchy based on Zipf's law and generalized allometric model was proposed (Chen, 1995; Chen, 2014b). Starting from the general system theory of Bertalanffy (1968), a general allometric growth equation can be derived as follows (Chen, 1995)

$$M(k) = \eta P(k)^p, \tag{21}$$

where $P(k)$ denotes the city size of the $k$th city, and $M(k)$ refers to some related response such as economic output, land use quantity, energy consumption, water consumption, and so on, $\eta$ is a proportionality coefficient, and $p$ is an allometric scaling exponent. Rewrite equation (2) as below

$$P_1 = U / \sum_{k=1}^{N} k^{-q}. \tag{22}$$

The sum of the above responses $M(k)$ is

$$M(p,q) = \sum_{r=1}^{n} M(r) = \eta \sum_{r=1}^{n} P(r)^p, \tag{23}$$

where $M(p, q)$ refers to the total response of a system of cities and towns. Substituting equations (1) and (22) into equation (23) yields the $q$-equation as follows

$$M(p,q) = \eta \sum_{r=1}^{n} (P_1 r^{-q})^p = K \sum_{r=1}^{N} (r^{-q} / \sum_{r=1}^{N} r^{-q})^p. \tag{24}$$

where $K=\eta U^p$. If $p>1$, $M(p, q)$ proved to be the increasing function of $q$, that is

$$dM(p,q)/dq > 0; \tag{25}$$

In contrast, if $p<1$, $M(p, q)$ proved to be the decreasing function of $q$, that is

$$dM(p,q)/dq < 0. \tag{26}$$

Empirical analyses shows that, if the response $M(k)$ represents urban economic output value, the scaling exponent $p>1$ (Chen, 1995; Chen and Liu, 1998; Chen and Zhou, 2003). This indicates *increasing return*. In contrast, if the response $M(k)$ represents urban land use quantity, the scaling exponent $p<1$ (Chen, 1995; Chen and Liu, 1998; Lee, 1989). This implies *agglomeration effect* and *scale economics*. What is more, if the response $M(k)$ represents urban energy and water consumption, the allometric scaling exponent $p>1$ in China (Chen, 1995; Chen and Liu, 1998). This means high carbon emissions, waste of resources, and environmental pollution. The inference is that the Zipf exponent, the $q$ value, cannot be too high or too low. If $q>1$, the total economic output of an urban system will be high and total urban land will be less. Meanwhile, the total quantity of energy and water consumption will also be high. If $q < 1$, the reverse is true: the total economic output of an urban system will be low and total urban land will be high. At the meantime, the total quantity of energy and water consumption will also be low. Zipf law indicates that cities bear no characteristic scale. In other words, cities have no typical size (Buchanan, 2000). However, the $q$-equation suggests that although cities do not have the best size, the urban system has the best size distribution (Figure 2). Where population is concerned, for small and medium-sized countries, the pure Zipf distribution represents the best size distribution of cities ($q$=1). However, for large population countries, this balance will lose, or the symmetry of the rank size distribution will break. The urbanization level will do not go up, or the Zipf scaling exponent will be less than 1, or even the Zipf distribution will have scaling breaking.

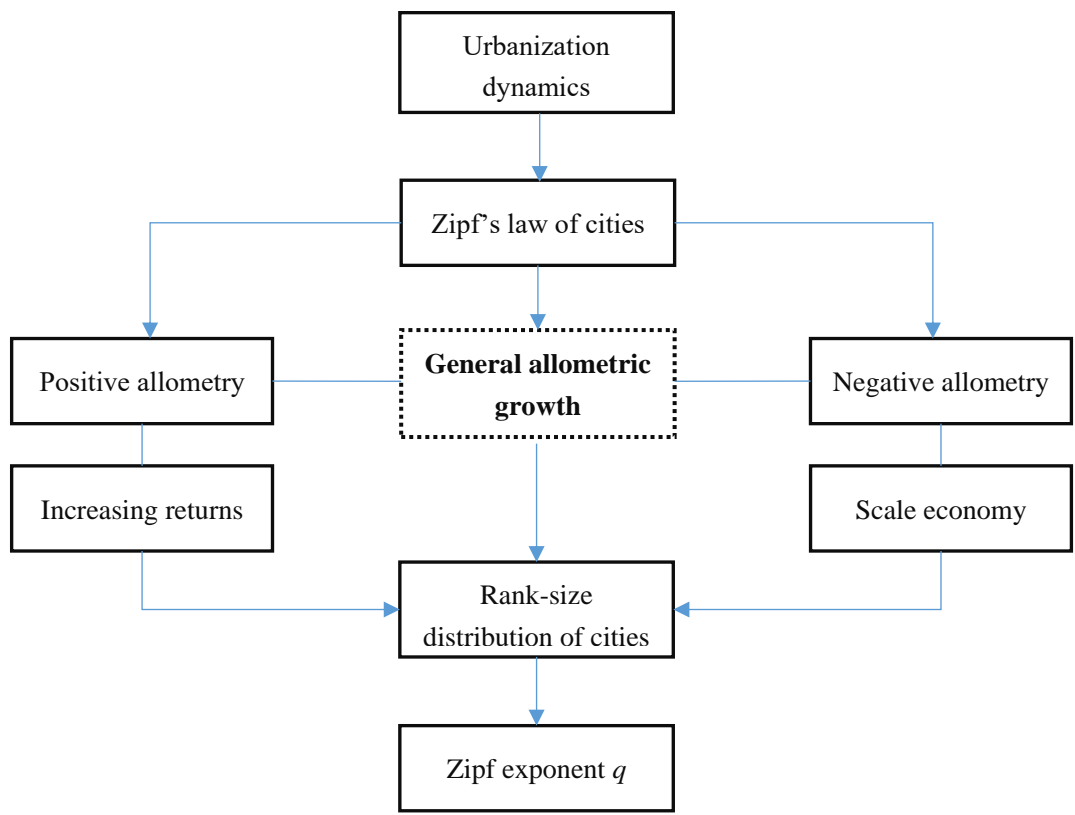

**Figure 2 The antecedents and consequences of the micro-level correlation between allometric growth and Zipf scaling exponent**

Note: At the micro level, urban elements involve population, land, infrastructure, and so on. The relationships between these elements follow allometric scaling law. There are positive and negative forces in these relationships. The push-pull effect of positive and negative forces makes the Zipf exponent $q$ tend to the appropriate value.

Associated with Zipf's law, the law of allometric growth is one of basic mathematical models in urban geography. Two Zipf's distributions support an allometric scaling relations (Chen, 2014b). The allometric growth law was introduced into urban studies early by Naroll and Bertalanffy (1956).

Because the allometric scaling exponent cannot be reasonably explained by Euclidean geometry, the related research once declined for a time. Due to introduction of fractal geometry, allometric studies on cities revived at the turn of the century (Batty and Longley, 1994; Chen, 1995; Chen and Xu, 1999; Lo, 2002). Based on equation (23), a number of interesting allometric scaling analyses have been made recent years (Arcaute *et al*, 2015; Bettencourt, 2013; Bettencourt *et al*, 2007; Chen, 1995; Lobo *et al*, 2013; Louf and Barthelemy, 2014a; Louf and Barthelemy, 2014b). A revealing finding is that the calculated values of the scaling exponent comes between 2/3 and 4/3 (Bettencourt, 2013). Some calculated values are less 2/3 (close to 1/2), while the other calculated values are greater than 4/3. The threshold value seems to be *p*=1, which forms a dividing line between two urban economic processes: increasing returns ($p>1$) and economies of scale ($p<1$) (Bettencourt *et al*, 2007). Arcaute *et al* (2015) found that the scaling exponent values of the allometric relation between patents and city population sizes relies on the cut-offs of city sizes. Louf and Barthelemy (2014a; 2014b) discovered that the scaling exponent values of the allometric relation between urban $CO_2$ emissions and city population sizes depend on the definition of urban area. A typical result is the allometric scaling exponent of the allometric relation between urban area and population. The expected value is about 0.85 (Chen, 2008a; Chen, 2010), and the empirical values are close to 0.85 (Louf and Barthelemy, 2014a; Chen and Xu, 1999). These studies not only lead to new achievements (Arcaute *et al*, 2015; Bettencourt, 2013; Bettencourt *et al*, 2007; Chen, 2014b; Lobo *et al*, 2013; Louf and Barthelemy, 2014a), but also to confusing problems (Arcaute *et al*, 2015; Chen and Lin, 2009; Louf and Barthelemy, 2014a; Louf and Barthelemy, 2014b). The puzzling problems may in turn result in new researching results. Bettencourt (2013) and his co-workers have develop new models of urban scaling, which may cause a new theory of city size in the future (Batty, 2013). Based on the results from Bettencourt (2013) and Bettencourt *et al* (2007), the *q*-equation can be further developed to explain the Zipf scaling exponent on city-size distribution and the dynamics of city development from a novel angle of view.

There are numerous papers on Zipf's law and numerous works on urbanization. The relationships between city size distribution and urbanization has been researched for many years. Compared with previous studies, the significant novelty of this work is as follows. First, now models and formulae are derived to reveal the relationships between urbanization level, city number, the population size of largest city, national population, and Zipf scaling exponent. Second, the spheres of application of pure Zipf law, general Zipf law, and the law of primate city are made clear. Third, new understanding about Zipf law and city size distribution are obtained from the mathematical reasoning and computational analyses. The main disadvantages of this study are as below. First, due to the lack of continuous time series data of cities and urbanization, dynamic analysis and verification cannot be carried out based on the mathematical models proposed in this paper. Second, due to the limitation of the length of the paper, the generalized models of rank-size distributions of cities have not been discussed for the time being. As indicated above, several scholars developed more than one types of general model for size distributions of cities (Ausloos and Cerqueti, 2016; Benguigui and Blumenfeld-Lieberthal, 2011). The problem will be solved in future studies.

## 5. Conclusions

This is a theoretical study on the relationships between the level of urbanization and the types of Zipf's city-size distribution. Although the problem is old, there are new discoveries and understandings on Zipf's law and urbanization in this paper. Previous studies on the size distribution of cities were majorly aimed at the countries with small population size. China, India, or even the future United States of America are different because of their large population. The models above-presented can be employed to answer the following questions: Why does China's city size distribution deviates from the pure Zipf's law ($q$<1), and why India's urbanization level does not reach the level of urban majority ($L$<50%)? And so on. Based on the theoretical derivations, computational analysis, and empirical evidence, the main conclusions can be reached as follows. (1) The pure Zipf size distribution of cities is suitable for median and small countries, not suitable for large populous countries such as China and India. If the city size distribution of a populous country follows the pure Zipf's law (scaling exponent $q$=1), the level of urbanization will be limited under 50%. If urbanization level goes up strongly, urban dynamics will force the Zipf scaling exponent $q$ to depart from 1 and become less than 1. Another possible case is that the scaling of rank-size distribution will break into two parts, and thus two scaling ranges will appear on a log-log plot for rank-size distribution. (2) Based on power-law distribution of cities, the Zipf scaling exponent is the control parameter, dominating the level of urbanization. The common Zipf model bears two parameters: proportionality coefficient and scaling exponent. The proportionality coefficient is a local parameter, indicating the population of the largest city. Its impact on the level of urbanization is limited and not significant. The scaling exponent is a global parameter, representing the characteristic parameter of city-size distribution. It significantly affects the level of urbanization through influencing the number of cities and towns in a country. The size of the largest city relies to a degree on the definition of cities in a country, and it can be properly determined by CCA. City definition does no influence the analytical conclusion of this study. (3) Besides the above two parameters, there is still two variables influencing urbanization level of a country, namely, the total national population and city number. For given total national population and Zipf exponent, the city number and the population of the largest city play an important part. The population of the primacy city is the characteristic value of total urban population. The number of cities can be reasonably determined by the scaling range of the rank-size distribution. If city-size distribution follows Zipf's law, changing city definition does not change the main inference for the relationships between rank-size scaling exponent and level of urbanization. (4) The spatial competition between city number and the population of largest city may result in primate distribution of city sizes. Under the condition that the population of a country is not too large, and the primary city participate in the evolution of the international urban network, the largest city may stand out from the rest and become very protruding in size. In this way, the largest city will form a shadow effect in the hierarchy and restrict the growth of other larger cities. Therefore, the local destruction of the city rank-size pattern appears. The primate distribution of city sizes is a local disturbance of the global size distribution of cities.

**Acknowledgements**

This research was sponsored by the National Natural Science Foundation of China (Grant No. 41671167). The support is gratefully acknowledged.